# Specific heat at low temperatures in quasiplanar molecular crystals: Where do glassy anomalies in minimally disordered crystals come from?

Daria Szewczyk[1,2,3], Manuel Moratalla[1,2,4], Grzegorz Chajewski[3], Jonathan F. Gebbia[5], Andrzej Jeżowski[3], Alexander I. Krivchikov[3,5,6], María Barrio[5], Josep Ll. Tamarit[5,*], Miguel A. Ramos[1,2,4,*]

[1]Laboratorio de Bajas Temperaturas, Departamento de Física de la Materia Condensada, Universidad Autónoma de Madrid, 28049 Madrid, Spain

[2]Instituto Nicolás Cabrera (INC), Universidad Autónoma de Madrid, 28049 Madrid, Spain

[3]Institute of Low Temperature and Structure Research PAS, 50-422 Wrocław, Poland

[4]Condensed Matter Physics Center (IFIMAC), Universidad Autónoma de Madrid, 28049 Madrid, Spain

[5]Grup de Caracterizació de Materials, Departament de Fisica, EEBE, and Barcelona Research Center in Multiscale Science and Engineering, Universitat Politècnica de Catalunya, 08019 Barcelona, Catalonia, Spain

[6]B. Verkin Institute for Low Temperature Physics and Engineering, NASU, 61103 Kharkiv, Ukraine

*Corresponding author. miguel.ramos@uam.es

*Corresponding author. josep.lluis.tamarit@upc.edu

**Abstract**

We present low-temperature specific heat ($C_p$) measurements of a monoclinic $P2_1/c$ crystal formed by quasiplanar molecules of tetrachloro-$m$-xylene. The dynamic disorder frozen at low-temperature of the asymmetric unit (formed by a half molecule) consists of reorientation around a three-fold-like axis perpendicular to the benzene ring. Such a minimal disorder gives rise to typical glassy anomalies, as a linear in contribution in $C_p$ ascribed to two-level systems and a broad maximum around 6.6 K in $C_p/T^3$ (the boson peak). We discuss these results in the framework of other quasiplanar molecular crystals with different accountable number of in-plane molecular orientations We find that the density of two-level systems does not correlate with the degree of orientational disorder. Rather, it is the molecular asymmetry that seems to play a relevant role in the thermal anomalies. Furthermore, we discuss the suggested correlation between the boson peak ($T_{BP}$) and Debye ($\Theta_D$) temperatures. We find that a linear correlation between $T_{BP}$ and $\Theta_D$ holds for many −but not all− structural glasses and strikingly holds even better for some disordered crystals, including our studied quasiplanar molecular crystals.

# I. INTRODUCTION

Since the middle of the last century, it is well known (although not so well understood) that structural glasses or amorphous solids present "anomalous" physical properties at low temperatures [1–4], which are very different from those found in crystals, that presumably follow Debye's theory [5–7]. Furthermore, during the last decades, it has been discovered that different crystals with some kind of orientational disorder ("orientational glasses") also show very similar glassy anomalies at low temperatures [3, 4, 8–16], and share many other dynamic properties with structural glasses. Even more recently, different types of crystals with minimal disorder within their crystalline unit lattices have also been found to exhibit such glassy behavior in their thermal, vibrational and dynamic properties at low temperatures [17–22].

Such "glassy anomalies" at low temperatures usually refer to a specific heat $C_p(T)$ much higher than that of crystals, with a quasi-linear dependence at very low temperatures followed by a wide maximum in $C_p/T^3$ usually around 3–10 K [1, 2, 15], as well as to a universally low thermal conductivity $\kappa(T)$ for all glassy materials that depends quadratically with temperature until reaching a plateau around 3–30 K followed by a subsequent increase at higher temperatures [1–3, 16]. In addition, the main low-temperature dielectric and acoustic properties of non-crystalline solids also universally show [2–4, 12, 23–25] very different behavior from that of their crystalline counterparts.

Below, say, 1 K, the aforementioned thermal, acoustic and dielectric properties exhibited by glassy matter were soon explained phenomenologically by the well-known Tunneling Model, proposed independently by Phillips [26] and by Anderson, Halperin and Varma [27]. For temperatures above 1 K, it is accepted that the excess of specific heat, acoustic and dielectric attenuation, and scattering of the acoustic vibrations acting as heat carriers in glasses, over those in crystals, are due to particular low-energy vibrational excitations produced by disorder and characterized by the so-called "boson peak", a ubiquitously observed maximum at low frequencies $\omega$ in the reduced density of vibrational states $g(\omega)/\omega^2$ obtained by various spectroscopy techniques [28–30]. However, the origin and microscopic nature of these excitations and the boson peak are still very controversial, with a large number of theories and models in this regard [4, 31–43].

What are the key ingredients that produce such ubiquitous glassy behavior associated with the presence of *tunneling states* or two-level systems (TLS), as well as a *boson peak* (BP) in the Debye-reduced vibrational density of states, remains an open question [44, 45] and a matter of vivid debate [4].

In this work, we present measurements of the specific heat in the range 0.15–25 K for crystals of tetrachloro-*m*-xylene (TCMX, see inset in Fig. 1), performed in three different experimental setups. Interestingly, a linear term in the specific heat $C_p$ is clearly observed below 1 K (ascribed to the presence of TLS), as well as a maximum in $C_p/T^3$ (the BP) at ≈6.6 K. We have determined the different coefficients (contributions) in the specific heat using the Soft-Potential Model (SPM) [31, 32, 46, 47] and compared them to those obtained in other molecular crystals with a similar quasiplanar molecular structure, such as pentachloronitrobenzene (PCNB), $C_6Cl_5NO_2$ [18] and *para*-chloronitrobenzene (*p*-CNB), $C_6H_4ClNO_2$ [17] –in the latter material also presenting new data at low temperature– to seek for possible correlations and clues concerning the physics behind.

Through an exhaustive knowledge of the crystal structure, we can obtain detailed information on which orientational states of the molecules are possible and how they manifest in these low-temperature metastable crystalline phases. This provides a key advantage over structural glasses, which have multiple sources of disorder. The quasiplanar molecule of hexa-substituted benzene, PCNB, is formed by a benzene ring decorated with a $NO_2$ group and five chlorine atoms. The room temperature phase, which remains metastable at 0 K, displays a layered structure of rhombohedral (space group $R\bar{3}$, $Z = 3$) planes in which the dynamics of the disorder involves in-plane reorientations about the axis normal to the benzene ring (i.e. six-fold-like axis) [18, 48, 49].

As far as *p*-CNB is concerned, the room temperature crystal structure (space group $P2_1/c$, $Z = 2$), which can be supercooled down to 0 K, exhibits two-fold orientational disorder of the $NO_2$ and Cl substituents, lying across an inversion centre (i.e. head-to-tail disorder concerning molecular orientation) [50–52].

The structure of TCMX has been determined by X-ray powder diffraction. The X-ray patterns obtained at 100 K were submitted to a Rietveld refinement with the rigid body constraints. Details for experimental and refined patterns can be found in Appendix A. The relevant result concerning the dynamic disorder of the $P2_1/c$ structure is that the asymmetric unit (formed by a half molecule) displays a three-fold-like axis perpendicular to the benzene ring, in such a way that the asymmetric unit can occupy three different in-plane orientations.

All in all, the different disordered phases for each of the three selected materials provide a set of low-temperature phases with an accountable number of (in-plane) molecular orientations: 2 for *p*-CNB, 3 for TCMX and 6 for PCNB. Disregarding the possible differences of fractional populations in the various orientations (because crystals are keeping the inversion symmetry) [53], one would expect that the evidenced differences in orientational disorder giving rise to multiple local potential minima would have a rational and scalable influence on low-temperature glassy properties. They will therefore be more amenable to a microscopic view of the underlying physics, which would help unravel the essential ingredients that produce glassy anomalies at low temperatures/energies in so many solid materials, both amorphous and (somewhat unexpectedly) also many crystalline ones.

In Section II, we describe the material and the experimental techniques used to measure the specific heat of TCMX. Additional characterization experiments, such as X-ray diffraction and Differential Scanning Calorimetry, are provided in Appendix A. In Section III, we present the specific heat results obtained for TCMX at low temperatures, followed by a discussion in Section IV which includes a comparison of our data with previously published and unpublished data on related PCNB and *p*-CNB quasiplanar crystals, as well as an exploration of possible correlations between the boson peak temperature and the Debye temperature. Complementary data plots and Soft-Potential Model fits are provided in Appendix B. A summary and conclusions of our work are given in Section V.

## II. EXPERIMENTAL

### A. TCMX material

TCMX (tetrachloro-*m*-xylene or 1,2,3,5-tetrachloro-4,6-dimethylbenzene, $C_8H_6Cl_4$, see inset in Fig. 1) was purchased from Dr. Ehrenstorfer GmbH with purity >99.5%. $M_w$ = 243.95 g/mol and melting temperature $T_m$ = 496 K, as confirmed by Differential Scanning Calorimetry, see Appendix A.

### B. Heat capacity experiments

Three different experimental setups were employed for performing heat capacity measurements, allowing to consistently cover a wide low-temperature range. This was also done to check reproducibility and ensure accuracy of the sought low-temperature specific-heat coefficients in these crucial cases where their values might be relatively low, especially the lowest-temperature linear coefficient that could be expected to be null as in ordinary crystals.

Specifically, in the Low-temperature Laboratory at Universidad Autónoma de Madrid, a homemade calorimetric system [54, 55] placed in a $^3$He cryostat, was employed. The calorimetric setup comprises a copper ring, which acts as a thermal source, with a resistor chip as a heater and a Cernox 1010 thermometer which are driven by a Lakeshore Temperature Controller 335. In this case, compacted crystallites of TCMX (*m* = 506.3 mg) were introduced into a thin-walled copper cell consisting of two concave plates, in a He atmosphere to avoid water absorption. The copper plates were pressed just at the same moment they were glued with the aim of improving the inner thermal contact between TCMX crystallites, as well as of them with the copper cell. The copper cell was then put on a sapphire disk with a 1000 Ω chip resistor as a heater and another $RuO_2$ chip as thermometer to monitor the temperature changes in the sample caused by the heater. The cell was previously introduced in a high-vacuum chamber, observing a null loss of mass after several hours maintaining a high vacuum in the setup, confirming a good vacuum seal. Once the measurements of the TCMX material inside the copper cell were finished, the empty addenda was carefully measured.

On the other hand, in Wrocław, two commercial Quantum Design PPMS platforms were used, one in a standard $^4$He configuration and another equipped with a $^3$He–$^4$He dilution refrigerator (DR). In this case, two pellets of masses *m*=7.1mg ($^4$He) and *m*=20.2mg (DR) were prepared using a mechanical press. The so-prepared samples were placed directly on the measuring tables of each system (so-called *pucks*) and secured with an apiezon grease. Before measuring the sample heat capacity, the contribution of the addenda was determined by measuring an empty puck with the apiezon grease.

## III. SPECIFIC HEAT RESULTS

Fig. 1 shows the specific heat of TCMX crystals, plotted in the usual Debye-reduced $C_p/T^3$ representation. Data obtained utilizing different experimental setups and methods show an excellent agreement, which supports their high reliability and accuracy. Over the cubic Debye contribution (horizontal dashed line), an upturn at the lowest temperatures (due to TLS) and a broad maximum (BP) typical of glasses are clearly observed. The extended dataset is shown in Appendix B (see Fig. 9).

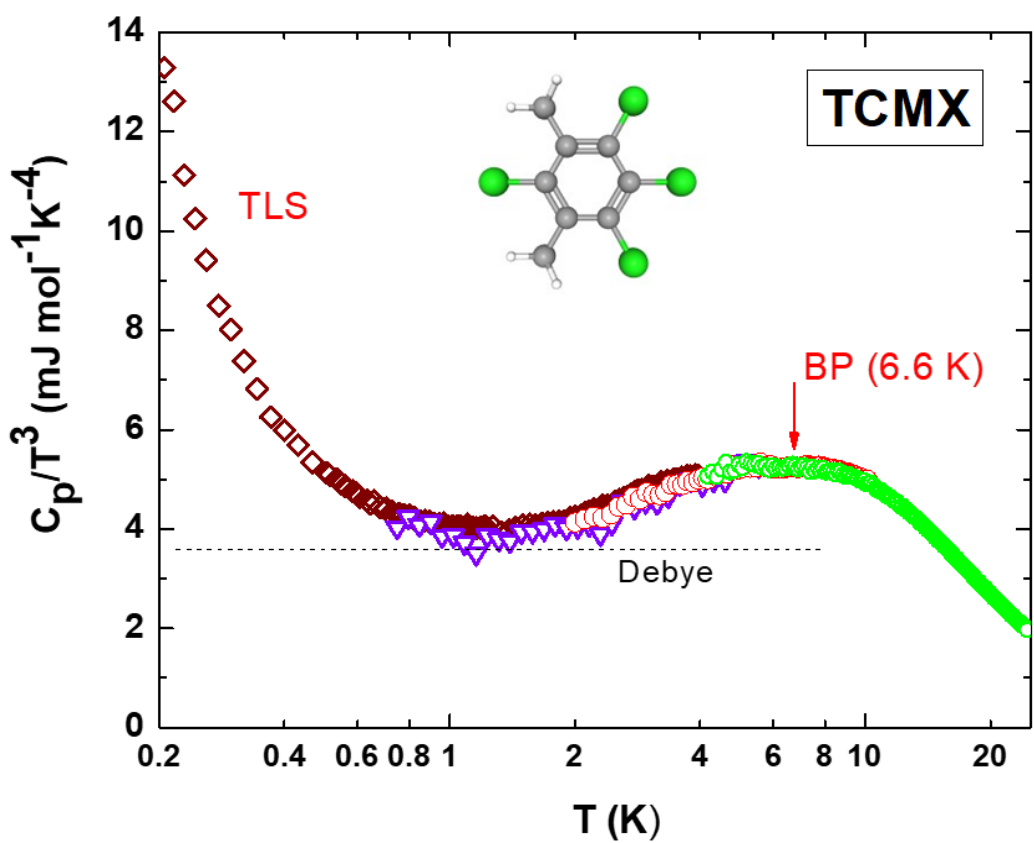

Figure 1. Specific heat of TCMX crystals in the Debye-reduced $C_p/T^3$ representation. The full set of experimental data points at even lower temperatures is shown in Appendix B. Different symbols correspond to data obtained in different experimental setups: DR (lozenges); $^3$He cryostat (down triangles); $^4$He PPMS (open circles, two different runs). TCMX molecule sketch: Grey, white and green balls represent C, H and Cl atoms, respectively. Horizontal dashed line: Debye contribution.

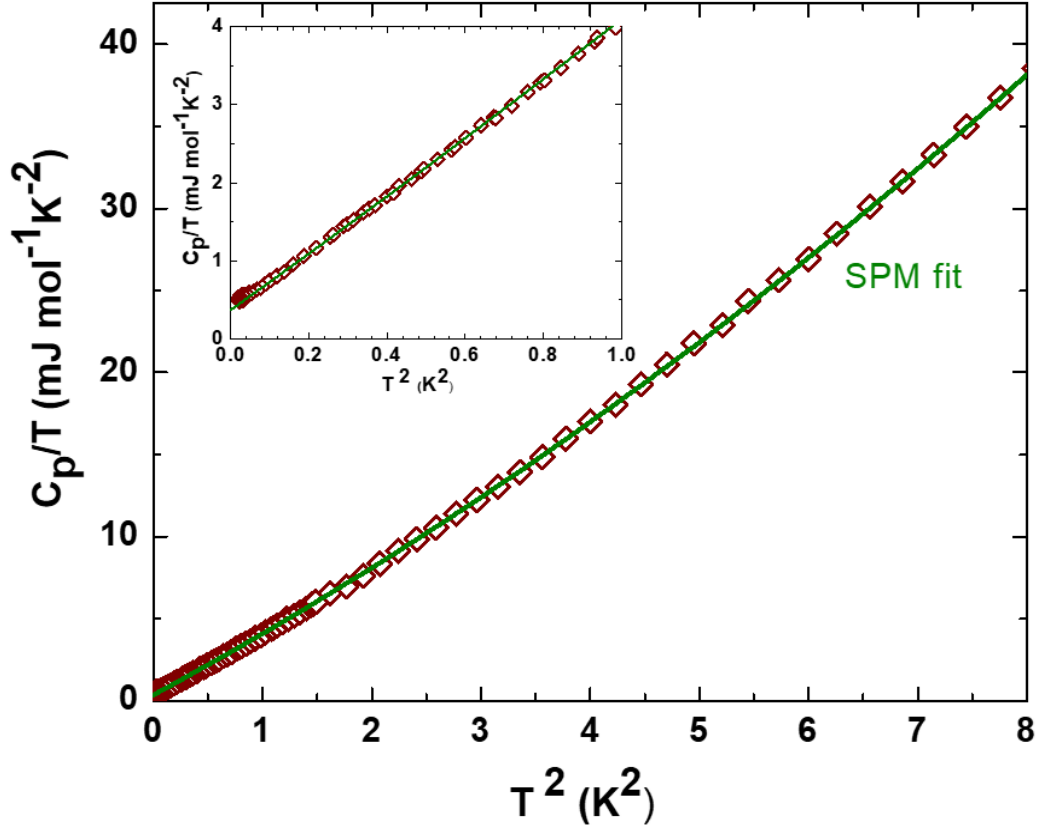

Figure 2. Specific heat of TCMX crystals at the lowest achieved temperatures, plotted in a $C_p/T$ vs $T^2$ representation. The solid line shows the fit to the Soft-Potential Model (see text). The inset amplifies data below 1 K.

Fig. 2 shows specific heat plotted as $C_p/T$ vs $T^2$ in the lowest temperature region, most relevant to determine the presence of a linear term attributed to TLS. The inset amplifies data below 1 K, evidencing the nonzero value of the linear term, given by the intercept with the ordinate axis. Also shown is a least-squares quadratic fit of the data to the Soft-Potential Model in the appropriate temperature range, given by $C_p = C_{TLS} T + C_D T^3 + C_{sm} T^5$ [46, 47], where $C_{TLS}$ stands for the TLS linear term contribution, $C_D$ is the classical Debye coefficient from acoustic phonons and $C_{sm}$ comes from the contribution of the low-frequency (soft) modes at the lower energy tail of the BP. The obtained coefficients are given in Table 1. It should be noted that with data obtained at such low temperatures (see inset of Fig. 2), the linear (TLS) and cubic (Debye) coefficients are practically independent of the model used for the analysis, whether with the quadratic fit of the SPM employed here or if a simple linear fit according to the Tunneling Model were used.

## IV. DISCUSSION

### A. Quasiplanar molecular crystals

The three different molecules under scrutiny, TCMX, PCNB and $p$-CNB, are depicted in Fig. 1 and in Figs. 3(a) and (b), respectively Our starting point is that the $p$-CNB molecule has two possible orientations between which a single molecule could rotate within its lattice site. The PCNB molecule presents up to six different orientations, whereas TCMX has three different orientations, though lacking the $NO_2$ group existing in the others.

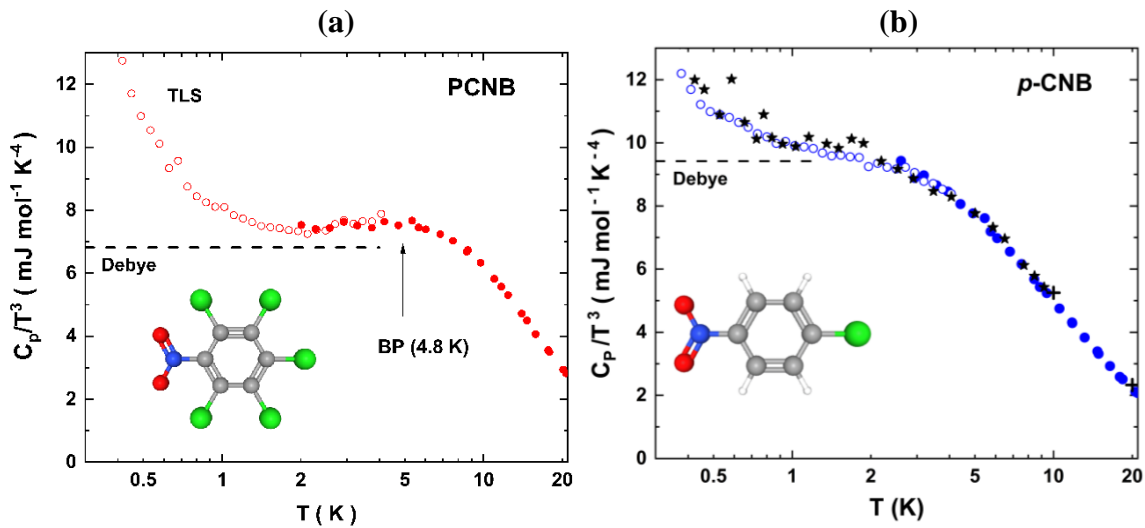

Figure 3. Specific heat Debye representation of (a) PCNB crystals, where different symbols indicate different experimental runs [18]; and (b) *p*-CNB crystal, showing both data taken from the literature (black stars [17]; black crosses [56]) and our new, unpublished data (open and solid blue circles, two different runs). The corresponding Debye contributions (cubic terms) are displayed by dashed lines and the position of the boson peak in PCNB is marked by an arrow. Molecules sketches: grey balls: C; white balls: H; green balls: Cl; blue balls: N; red balls: O.

In Fig. 3 we present both published and new specific-heat data for the other two quasiplanar molecular crystals with which we seek to compare. Fig. 3(a) shows the Debye-reduced $C_p/T^3$ plot previously obtained for the PCNB crystal [18], with a shallow BP at 4.8 K and a significant upturn over the Debye level at lower temperatures, signalling the presence of TLS. Fig. 3(b) contains $C_p/T^3$ data for *p*-CNB, both previously published data from the literature [17, 56] and our own recent measurements. As can be seen, it also shows a clear upturn at the lowest temperatures ascribed to TLS, but not a BP. The absence of a BP for the *p*-CNB disordered crystalline phase resembles the behavior found in other glasses and molecular crystals [57−60].

We have performed fitting of the SPM for low-temperature data in the $C_p/T$ vs $T^2$ representation for both PCNB and *p*-CNB crystals with glassy behavior. These fits are shown in Appendix B (see Figs. 10 and 11), and the coefficients obtained from them are gathered in Table 1, together with other useful data. Let us note that the negative fifth-power coefficient $C_{sm}$ for *p*-CNB obtained from the fit is obviously a consequence of the depression of $C_p/T^3$ in the BP temperature domain.

The first clear conclusion from the data in Table 1 is that there exists no direct correlation between the glassy properties and the available number of orientational states (orientational potential minima) in each material. The density of TLS (either per molecule or per atom) is minimal for the middle case of TCMX with 3 orientational states. Also for this central case of TCMX, the BP occurs at a higher temperature (6.6 K) and has the best defined (the strongest) relative maximum over the Debye background ($\Delta C_p/T^3$). There is also no correlation with the number of optical modes (see number of atoms for each molecule in Table 1 or with the number of molecules $Z$ in the primitive unit cell in Table 2).

With respect to structural symmetry, TCMX and PCNB [49] exhibit 2D stacked structures where molecules are π-stacked in columns, unlike the structure of the disordered phase of p-CNB [52], despite their closely related molecular symmetry. Such π-π interactions are known to play a significant role in organic systems, facilitating pathways for charge carriers and electrons, which in turn highlight an asymmetry in thermal conductivity between in-plane and out-of-plane directions of the 2D structure or can lead to negative thermal expansion [61–64]. Additionally, the absence of a low-temperature heat capacity peak has been observed in other organic crystals and biological systems [61, 65–68] that contain aromatic molecular units with π−π stacking, where flexural modes may arise [69]. It has been speculated that in such systems, the 2D structural features combined with π-π interactions could be the reason behind the absence of the boson peak (BP). Our work demonstrates that disordered molecular systems with specific 2D stacking, like TCMX and PCNB, still exhibit the BP, while *p*-CNB does not. Further studies on molecular systems composed of aromatic entities with and without π-π interactions should be conducted to clarify their effects on low-temperature thermal properties. The presence of flexural modes, giving rise to the long-wavelength ZA phonons predicted long ago by elasticity theory [70], is critical in determining the density of states and, therefore, the thermal properties of these materials.

Moreover, such studies could enhance the understanding of amorphous materials in general, and of amorphous drug systems in particular. For these cases, characterizing molecular mobility in the amorphous state via terahertz spectroscopy (around the BP region) is particularly important for improving drug delivery bioavailability [71].

Table 1. List of some material data for the three studied substances, as well as corresponding experimental results and specific-heat coefficients obtained from the SPM equation $C_p = C_{TLS} T + C_D T^3 + C_{sm} T^5$. SPM coefficients are given both per unit mole and per gram-atom, i.e. per mol of atoms instead of molecules. $\Delta C_p/T^3$@BP stands for the excess specific heat over the Debye level evaluated at the boson peak temperature $T_{BP}$. For the case of *p*-CNB, not exhibiting a BP but just a shoulder at about 4 K, corresponding data are given in parenthesis.

|  | *p*-CNB | TCMX | PCNB |
|---|---|---|---|
| **No. of orientational states** | 2 | 3 | 6 |
| **$C_{TLS}$ (mJ/mol·K$^2$)** | 0.507 ± 0.081 | 0.371 ± 0.015 | 1.06 ± 0.11 |
| **$C_D$ (mJ/mol·K$^4$)** | 9.42 ± 0.08 | 3.59 ± 0.012 | 6.82 ± 0.08 |
| **$C_{sm}$ (mJ/mol·K$^6$)** | −0.043 ± 0.012 | 0.142 ± 0.002 | 0.072 ± 0.009 |
| $M_w$ (g/mol) | 157.55 | 243.95 | 295.32 |
| $\alpha$ (atoms/molecule) | 14 | 18 | 14 |
| Density @100 K (g/cm$^3$) | 1.584 | 1.800 | 2.043 |
| $C_{TLS}$ ($\mu$J/g-at K$^2$) | 36.2 ± 5.8 | 20.6 ± 0.8 | 75.7 ± 7.9 |
| $C_D$ ($\mu$J/g-at K$^4$) | 673 ± 6 | 199 ± 0.7 | 487 ± 6 |
| $C_{sm}$ ($\mu$J/g-at K$^6$) | −3.1 ± 0.9 | 7.89 ± 0.11 | 5.14 ± 0.64 |
| $T_{BP}$ (K) | (≈ 4) | 6.6 | 4.8 |
| $C_p/T^3$ @BP (mJ/molK$^4$) | (8) | 5.3 | 8 |
| $\Delta C_p/T^3$ @BP (mJ/molK$^4$) | - | 1.7 | 1.2 |

Therefore, trying to correlate the TLS contribution in disordered solids at low temperatures or the position and magnitude of the BP with the degree of orientational disorder seems too simplistic. Rather, our findings suggest that these low-frequency glassy excitations are not just non-interacting, localized excitations, but the result of coupling or hybridizing ubiquitous "glassy" defects (likely with participation ratios of tens of particles) with the usual lattice vibrations (phonons). This general idea would fit perfectly with some models or theories already proposed for glasses, both in the TLS range [44, 45] and also for the BP feature, with a vibrational density of states universally following $g(\omega) \propto \omega^4$ for $\omega \to 0$ [31, 32, 34, 35, 41-43, 46]. The relevance of hybridization between quasilocalized optical modes and acoustic phonons had already been emphasized in halomethane crystals with glassy behavior [19].

The energy barrier between the different molecular orientations in the crystal lattice sites, energetically equivalent due to the symmetry (two-fold-like axis for *p*-CNB, six-fold-like axis for PCNB and three-fold-like axis for TCMX) should be energetically favoured by the great similarity of the van der Waals radii of the chlorine atom and the methyl group compared to the nitro group.

Nonetheless, as far as the BP is concerned, we have unexpectedly found an interesting correlation between the BP temperature $T_{BP}$ in the studied quasiplanar crystals and the corresponding Debye temperatures $\Theta_D$. In Table 2, we present the Debye temperature of *p*-CNB, TCMX and PCNB, obtained from the cubic coefficients of the corresponding specific heat as described above, as well as for some other recently reported glassy crystals formed by quasiplanar molecules. Here $\Theta_D$ has been defined as usual, regarding the total number of *atoms* per unit volume, which implies considering the whole of phonon branches contributing to the vibrational density of states, either acoustic or optical. If $\alpha$ is the number of atoms per molecule, $\Theta_D$ is obtained from the Debye coefficient by

$$C_D = \frac{(1944 \cdot \alpha)}{\Theta_D^3} \text{J/mol} \cdot \text{K} \tag{1}$$

Table 2. Debye temperatures $\Theta_D$, boson peak temperatures $T_{BP}$, and their ratio, for the studied group of quasiplanar molecular crystals of benzene derivatives, as well as for a group of bromine-benzophenones, where $s$ stands for the stable crystal phase and $m$ stands for a metastable crystal phase [20]. The values for the number of atoms per molecule ($\alpha$) and of molecules in the primitive cell Z are also given.

|  | $\alpha$ | $Z$ | $\Theta_D$ (K) | $T_{BP}$ (K) | $\Theta_D/T_{BP}$ |
|---|---|---|---|---|---|
| **p-CNB** | 14 | 2 | 142.4 | ($\approx 4$) | ($\approx 35.6$) |
| **TCMX** | 18 | 2 | 213.6 | 6.6 | 32.4 |
| **PCNB** | 14 | 3 | 158.6 | 4.8 | 33.0 |
| **BZP** (*s*) | 24 | 4 | 248 | 7.1 | 34.9 |
| **2BrBZP** (*s*) | 24 | 4 | 242 | 7.6 | 31.8 |
| **2BrBZP** (*m*) | 24 | 4 | 258.5 | 7.2 | 35.9 |
| **3BrBZP** (*s*) | 24 | 8 | 227 | 7.6 | 29.9 |
| **4BrBZP** (*s*) | 24 | 4 | 239.5 | 6.9 | 34.7 |
| **4BrBZP** (*m*) | 24 | 2 | 233 | 6.7 | 34.7 |

As can be seen in Table 2, in the three studied (minimally disordered) crystals, the Debye temperatures are near 35 times the BP temperature, $\Theta_D \approx 35 T_{BP}$. Remarkably, this is the prediction proposed by Granato from his interstitialcy model [72–74] calculated in its simplest approximation, $T_{BP} \approx \Theta_D/35$, after some guesses about the numerical coefficients of typical materials. This prediction was reasonably fulfilled in some groups of structural glasses [75–77] but not in others [77]. It is however very striking that this precise correlation is indeed found for our case of quasiplanar crystals with a small amount of orientational disorder. Motivated by this finding, we have included in Table 2 previously published data for the BP and Debye temperatures (taking into account the number $\alpha$ of atoms per molecule) of a set of bromine-benzophenone (Br-BZP) crystals [20] for which glassy properties were found. This set comprises single crystals of benzophenone ($C_{13}H_{10}O$, BZP) and both stable and metastable ordered crystalline polymorphs of several bromine derivative isomers: 2-BrBZP, 3-BrBZP, and 4-BrBZP, that is 2-, 3-, and 4- bromobenzophenone, respectively ($C_{13}H_9OBr$ in all cases). These isomers just differ in the position of the Br atom in one of the phenyl rings.

We plot in Fig. 4 the values of $T_{BP}$ vs $\Theta_D$ for all the crystals from Table 2. The abovementioned Granato's linear correlation $T_{BP} \approx \Theta_D/35$ is also shown by a red solid line. The unexpected agreement is certainly very good.

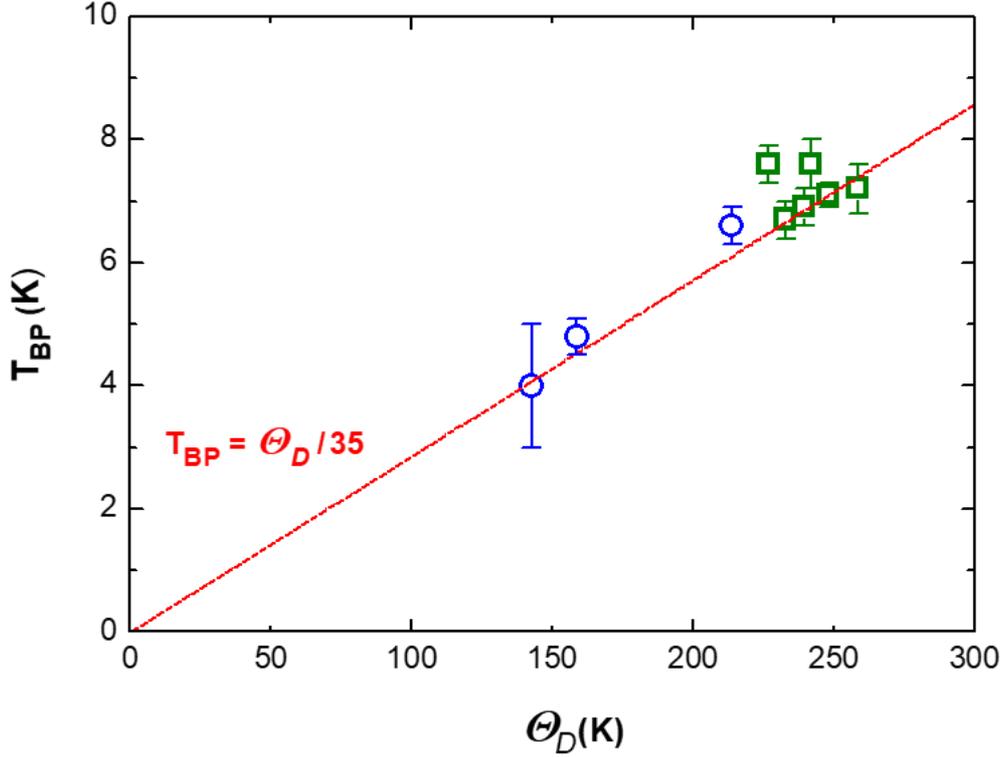

Figure 4. Relation of $T_{BP}$ vs $\Theta_D$ for all the crystals from Table 2: quasiplanar molecular crystals of benzene derivatives (circles) and bromine-benzophenones (squares). The Granato's linear correlation $T_{BP} \approx \Theta_D/35$ is shown by a solid line.

Therefore, not only our family of benzene-derivative molecular crystals but also that of bromine-benzophenone molecular crystals (regardless of being stable or metastable polymorphs) exhibit virtually within experimental error the "Granato's prediction" $\Theta_D \approx 35 T_{BP}$, following it even better than some structural glasses.

Granato's interstitialcy theory introduces concepts related to shear susceptibility and vibrational entropy [74], emphasizing their significance in understanding low-frequency vibrational modes in various materials. The theory predicts that a high shear susceptibility (defined as $-d \ln G/dc$, where $G$ is the shear modulus and $c$ the interstitialcy concentration) enhances the contribution of low-frequency vibrational modes, which can be particularly relevant for soft materials, such as organic molecular crystals. However, Granato's discussion primarily focused on comparing the properties of liquids, glasses, and crystals near melting and glass transition points. In the glass state at low temperatures, the interstitialcy concentration becomes effectively frozen, and thus the concept of shear susceptibility does not yield specific predictions for low-temperature properties of glasses that could be compared to experimental data. Granato's theory does predict that the resonant modes of the postulated interstitialcies, treated as Einstein modes, would contribute to the specific heat with a peak in $C_p/T^3$, roughly estimated to be around 1/35 of the Debye temperature. Additionally, the theory predicts that the Debye temperature of the glass is lower than that of the corresponding

crystal—a straightforward result, as the elastic properties of crystals are typically higher than those of their glassy counterparts.

In general, the interstitialcy theory could prove useful for systems where vibrational entropy is associated with internal defects or weak intermolecular interactions, which introduce additional degrees of freedom. In molecular crystals with π-π stacking, these characteristics can enhance soft vibrational modes by increasing the system's deformability under external stress. According to Granato's theory, defects or structural anisotropies may soften the shear modulus, thereby increasing susceptibility to low-frequency acoustic modes. This softening may ultimately influence low-temperature vibrational and thermal properties, such as the boson peaks observed in heat capacity measurements.

### B. Glasses and disordered crystals

Here we extend the discussion presented above to include many more disordered crystals and structural glasses whose needed data have been reported in the literature.

Table 3. Different values of the specific heat in the Debye-reduced representation, corresponding Debye temperatures $\Theta_D$, boson peak temperatures $T_{BP}$, and their ratio, for different crystals reported in the literature, including our data presented in the main text. Bromine-benzophenones (BZP) data were taken from Ref. [20]; thiophene data from Ref. [21]; cyclohexanol data from Ref. [8]; adamantanes data from Ref. [22]; TPD data from Ref. [78]. The rest of data were taken from Ref. [15]. The values for the number of atoms per molecule ($\alpha$) are also given in the table.

| | $\alpha$ | $(C_p/T^3)_{max}$ (mJmol$^{-1}$K$^{-4}$) | $(C_p/T^3)_{Debye}$ (mJmol$^{-1}$K$^{-4}$) | $\Theta_D$ (K) | $T_{BP}$ (K) | $\Theta_D/T_{BP}$ |
|---|---|---|---|---|---|---|
| **p-CNB** | 14 | (8) | 9.42 | 142.4 | ($\approx 4$) | ($\approx 35.6$) |
| **TCMX** | 18 | 5.3 | 3.59 | 213.6 | 6.6 | 32.4 |
| **PCNB** | 14 | 8 | 6.82 | 158.6 | 4.8 | 33.0 |
| **BZP (s)** | 24 | 4.1 | 3.06 | 248 | 7.1 | 34.9 |
| **2BrBZP (s)** | 24 | 6.46 | 3.3 | 242 | 7.6 | 31.8 |
| **2BrBZP (m)** | 24 | 5.11 | 2.7 | 258.5 | 7.2 | 35.9 |
| **3BrBZP (s)** | 24 | 6.32 | 4 | 227 | 7.6 | 29.9 |
| **4BrBZP (s)** | 24 | 6.42 | 3.4 | 239.5 | 6.9 | 34.7 |
| **4BrBZP (m)** | 24 | 6.84 | 3.7 | 233 | 6.7 | 34.7 |
| **n-thiophene (C$_4$H$_4$S)** | 9 | 2.34 | 1.23 | 242.3 | 8.4 | 28.8 |
| **d-thiophene (C$_4$D$_4$S)** | 9 | 2.52 | 1.31 | 237.3 | 8.6 | 27.6 |
| **cyclohexanol** | 19 | 3.79 | 2.58 | 242.8 | 4.49 | 54.1 |
| **1-fluoro-adamantane** | 26 | 2.62 | 1.25 | 343.2 | 10.8 | 31.8 |
| **2-adamantanone** | 25 | 1.74 | 1 | 364.9 | 12.2 | 29.9 |
| **cyano-adamantane** | 27 | 4.01 | 3.024 | 258.9 | 7.1 | 36.5 |
| **CBr$_2$Cl$_2$** | 5 | 17 | 5 | 124.8 | 7.5 | 16.6 |
| **CBrCl$_3$** | 5 | 8 | 2.3 | 161.7 | 7.7 | 21.0 |
| **CCl$_4$** | 5 | 6 | 1.81 | 175.1 | 9.2 | 19.0 |
| **H-ethanol: ODC** | 9 | 2.1 | 1.45 | 229.4 | 6.8 | 33.7 |
| **D-ethanol: ODC** | 9 | 2.6 | 1.72 | 216.7 | 6.4 | 33.9 |
| **TPD** | 72 | 19.6 | 14.6 | 212.4 | 4 | 53.1 |
| **Freon 112** | 8 | 11 | 4.43 | 152.0 | 4.5 | 33.8 |
| **Freon 113** | 8 | 11.5 | 3.81 | 159.8 | 5 | 32.0 |
| **ThBr$_4$** | 5 | 35 | 8 | 106.7 | 9.2 | 11.6 |
| **(NaCN)$_{0.81}$(KCN)$_{0.19}$** | 3 | 0.78 | 0.586 | 215.1 | 4.8 | 44.8 |
| **(NaCN)$_{0.41}$(KCN)$_{0.59}$** | 3 | 1.05 | 0.938 | 183.9 | 4.5 | 40.9 |
| **(NaCN)$_{0.75}$(KCN)$_{0.25}$** | 2.25 | 2.82 | 1.41 | 145.8 | 4 | 36.5 |
| **(NaCN)$_{0.30}$(KCN)$_{0.70}$** | 2.7 | 2.42 | 1.21 | 163.1 | 3 | 54.4 |

Table 4. Different values of the specific heat in the Debye-reduced representation, corresponding Debye temperatures $\Theta_D$, boson peak temperatures $T_{BP}$, and their ratio, for different structural glasses reported in the literature. Data were taken from Ref. [15] and references therein, except TPD data taken from Ref. [78]. The values for the number of atoms per molecule ($\alpha$) are also given in the table.

| | $\alpha$ | $(C_p/T^3)_{max}$ (mJmol$^{-1}$K$^{-4}$) | $(C_p/T^3)_{Debye}$ (mJmol$^{-1}$K$^{-4}$) | $\Theta_D$ (K) | $T_{BP}$ (K) | $\Theta_D/T_{BP}$ |
|---|---|---|---|---|---|---|
| **SiO$_2$** | 3 | 0.24 | 0.0462 | 501.6 | 10 | 50.2 |
| **GeO$_2$** | 3 | 0.315 | 0.205 | 305.3 | 8.4 | 36.3 |
| **B$_2$O$_3$ (quenched)** | 5 | 1.3 | 0.548 | 260.8 | 5.2 | 50.2 |
| **B$_2$O$_3$ (annealed)** | 5 | 1.2 | 0.465 | 275.5 | 5.6 | 49.2 |
| **(B$_2$O$_3$)$_{0.99}$ (Na$_2$O)$_{0.01}$** | 4.98 | 1.05 | 0.486 | 271.1 | 5.5 | 49.3 |
| **(B$_2$O$_3$)$_{0.94}$ (Na$_2$O)$_{0.06}$** | 4.88 | 0.7 | 0.254 | 334.3 | 7.8 | 42.9 |
| **(B$_2$O$_3$)$_{0.84}$ (Na$_2$O)$_{0.16}$** | 4.68 | 0.35 | 0.146 | 396.5 | 11.2 | 35.4 |
| **(B$_2$O$_3$)$_{0.75}$ (Na$_2$O)$_{0.25}$** | 4.5 | 0.25 | 0.098 | 446.9 | 11.9 | 37.6 |
| **As$_2$S$_3$** | 5 | 3.9 | 1.94 | 171.1 | 4.9 | 34.9 |
| **Se** | 1 | 2.5 | 1.4 | 111.6 | 3.1 | 36.0 |
| **glycerol** | 14 | 1.4 | 0.855 | 316.9 | 8.7 | 36.4 |
| **toluene** | 15 | 6.4 | 4.7 | 183.8 | 4.5 | 40.8 |
| **H-ethanol** | 9 | 2.4 | 1.55 | 224.3 | 6.1 | 36.8 |
| **D-ethanol** | 9 | 2.8 | 1.8 | 213.4 | 6 | 35.6 |
| **1-propanol** | 12 | 2.7 | 1.77 | 236.2 | 6.7 | 35.3 |
| **2-propanol** | 12 | 3.6 | 2.54 | 209.4 | 5 | 41.9 |
| **n-butanol** | 15 | 3.1 | 1.74 | 255.9 | 5.4 | 47.4 |
| **sec-butanol** | 15 | 4.5 | 1.82 | 252.1 | 4.8 | 52.5 |
| **isobutanol** | 15 | 5.8 | 2 | 244.3 | 4.8 | 50.9 |
| **TPD ordinary** | 72 | 30.5 | 17.6 | 199.6 | 2.8 | 71.3 |
| **TPD ultrastable** | 72 | 24.8 | 18.1 | 197.8 | 3.5 | 56.5 |
| **PMMA** | 15 | 3.3 | 1.93 | 247.2 | 3.6 | 68.7 |
| **PS** | 16 | 6.1 | 3.64 | 204.4 | 3 | 68.1 |
| **PB** | 10 | 2.2 | 1.08 | 262.1 | 5.1 | 51.4 |

In Table 3 we present literature data for 28 different crystals exhibiting some kind of "glassy behavior", whereas in Table 4 we present similar data for 24 different structural glasses. Then, in Fig. 5 $T_{BP}$ is plotted against $\Theta_D$ for those crystals and glasses, (upper and lower panel, respectively), in a log-log scale to facilitate their visualization. A similar plot −although in a linear scale− was employed by Carini *et al.* [75] and by Zhu and Chen [76] for structural glasses, finding $\Theta_D \approx 38-40 T_{BP}$.

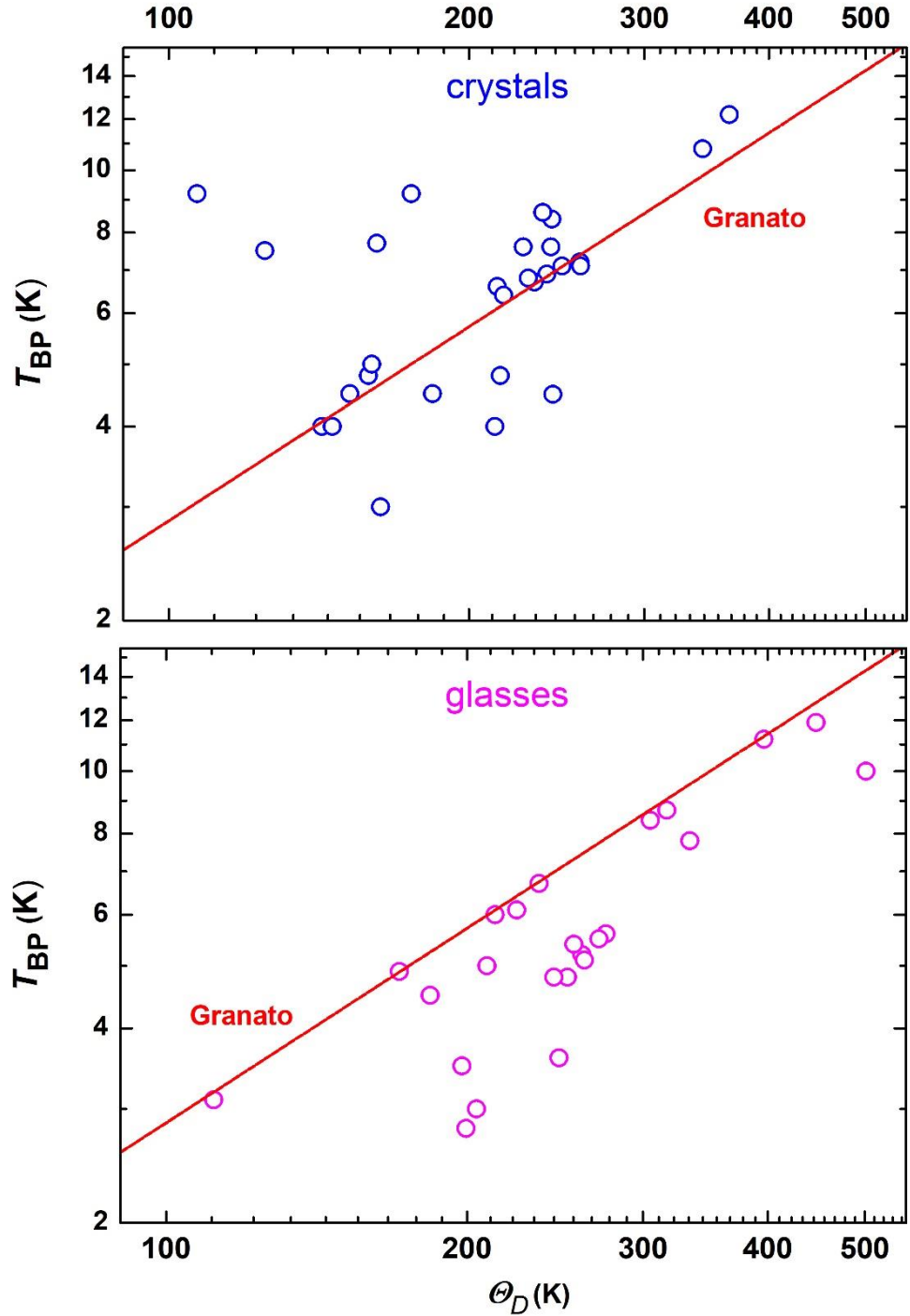

Fig. 5. Boson peak temperature ($T_{BP}$) plotted against Debye temperature ($\Theta_D$) for crystals (Table 3) and for glasses (Table 4), in a log-log scale. Granato's prediction $T_{BP} \approx \Theta_D/35$ is indicated by a red line.

Furthermore, in Fig. 6 corresponding histograms of the ratio $\Theta_D/T_{BP}$ are shown. As can be seen in all these figures, in (more-or-less disordered) crystals, the Granato's ratio $\Theta_D/T_{BP} \approx 35$ coincides well with the average value. Nevertheless, in structural glasses this ratio is indeed observed in many cases (in agreement with abovementioned statements in the literature), but a more thorough inspection indicates that it is rather a sort of lower bound. Many glasses exhibit $\Theta_D/T_{BP}$ ratios well above 35, but none of them clearly below it, in contrast to the case of crystals.

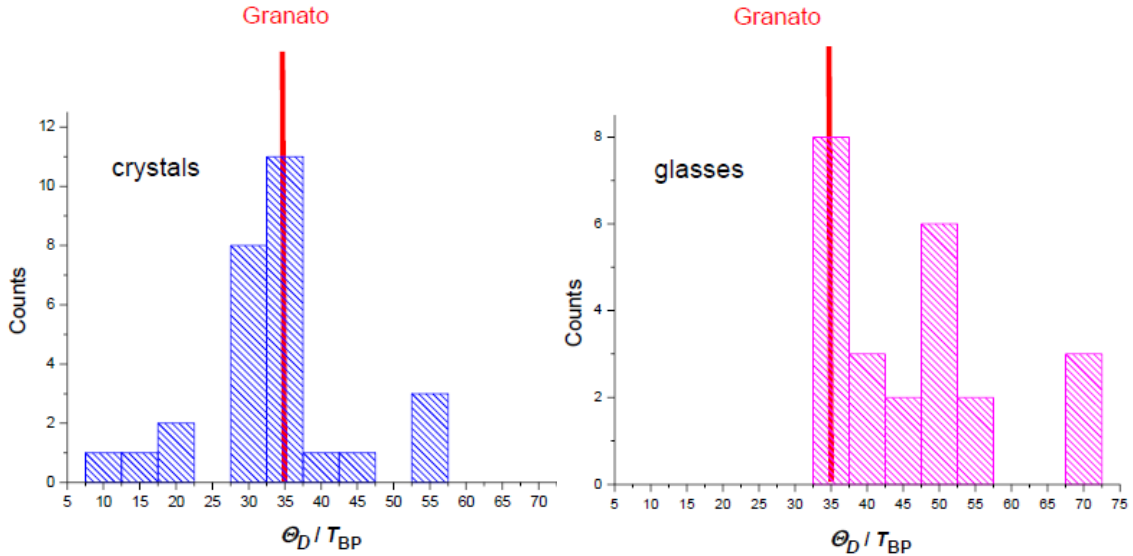

Fig. 6. Corresponding histograms of the ratio $\Theta_D/T_{BP}$ for crystals (left panel) and for glasses (right panel). The value expected from Granato's prediction $\Theta_D/T_{BP} \approx 35$ is marked by a solid bar.

It is important to stress that this quantitative correlation was obtained [61, 62] assuming that all interstitial resonance modes are the same within the Einstein model (which definitely is not the case observed in vibrational spectroscopy) and taking a simple cubic metallic crystal as a reference. As a matter of fact, it is hard to believe that such a simplified model could account for the dynamics neither of glasses nor of disordered crystals. In our view, these findings rather lend support to the existence of a strong link between the BP feature in the reduced vibrational density of states of disordered solids and the Debye temperature/frequency associated with the whole of the lattice dynamics of the material. In other words, the BP marks the threshold of interacting/hybridizing quasilocalized vibrations and extended phonons. This picture of the BP is essentially what is supported by several models, including the vibrational instability model [34, 35] built upon the Soft-Potential Model [32, 46], or the results of some computer simulations in glasses [41-43, 79, 80]. What is less clear is why their correlation ratio is so constant in many systems.

## V. CONCLUSION

We have performed low-temperature specific heat measurements in TCMX crystals, with a high degree of accuracy and reproducibility. We have found that quasiplanar molecular crystals of TCMX ($P2_1/c$) with a minimal disorder exhibit a boson peak at ≈ 6.6 K, as well as a density of two-level systems also typical of glasses. Compared to other similar molecular crystals (benzene derivatives) also with in-plane orientational disorder, TCMX shows up a similar BP as in PCNB, whereas the BP is absent in *p*-

CNB. However, all three orientationally-disordered crystals display the TLS fingerprint.

A bit surprisingly, the glassy features (TLS and BP) do not appear to correlate at all with the number of available orientations in the molecule, i.e. with the disorder of the entities forming the system. Tentatively, we attribute the observed differences to a more "symmetrical" molecule of TCMX lacking the $NO_2$ group, with only Cl and $CH_3$ substitutions, which have very similar van der Waals volumes. Hence the effective density of TLS would be influenced by the global distortion of the medium-range order in the solid lattice, rather than by isolated, non-interacting tunneling defects.

On the other hand, a striking correlation between the BP temperature and the Debye temperature is found, in agreement with Granato's prediction from the interstitialcy theory, $\Theta_D \approx 35 T_{BP}$. Moreover, this relation is fulfilled not only for these three benzene derivatives, but also for a set of bromine-benzophenone molecular crystals, not noticed until now, and also for many other (but not all) glasses and disordered crystals. We interpret these findings as a demonstration of a direct relation between the ubiquitous BP feature in non-fully-ordered crystals and their corresponding vibrational spectrum, the BP emerging as the central point of the hybridization of quasilocalized vibrations and extended phonons. Why the ratio unexpectedly tends to be the factor 35 suggested by Granato from his simple interstitialcy model in many more-or-less disordered crystals and in some fully ordered crystals (matching its average value), and in many structural glasses (though seemingly being here only a lower bound) remains an open question.

## ACKNOWLEDGEMENTS

D.S., M.M. and M.A.R. are grateful to J. M. Castilla for his valuable help with experiments. D.S. has been funded by the Bekker Programme of the Polish National Agency for Academic Exchange (Grant number BPN/BEK/ 2021/1/00091). M.M. and M.A.R. acknowledge financial support from MCIN/AEI/10.13039/501100011033 by the grant PID2021-127498NB-I00 and also through the "María de Maeztu" Program for Units of Excellence in R&D (CEX2018-000805-M). J.F.G., M.B. and J.Ll.T. acknowledge financial support from MCIN/AEI under project PID2023-146623NB-I00 and from Generalitat de Catalunya (2021SGR-00343). A.J. was supported by the Polish National Science Center (Grant No. 2022/45/B/ST3/02326) and A.I.K. was partly supported by the National Research Foundation of Ukraine (Grant 2023.03/0012). We also acknoweledge the Spanish Ministry of Universities for supporting A.I.K. through the Plan de Acción Universidad-Refugio.

## APPENDIX A: CHARACTERIZATION OF TCMX CRYSTALS

As shown in Fig. 7, the analysis of the X-ray diffraction pattern for TCMX crystalline powder at 100 K revealed a monoclinic structure ($P2_1/c$ space group), with the following lattice parameters: $a = 8.0565(9)$ Å; $b = 3.7959(5)$ Å; $c = 17.0031(18)$ Å; $\beta = 120.048(5)°$.

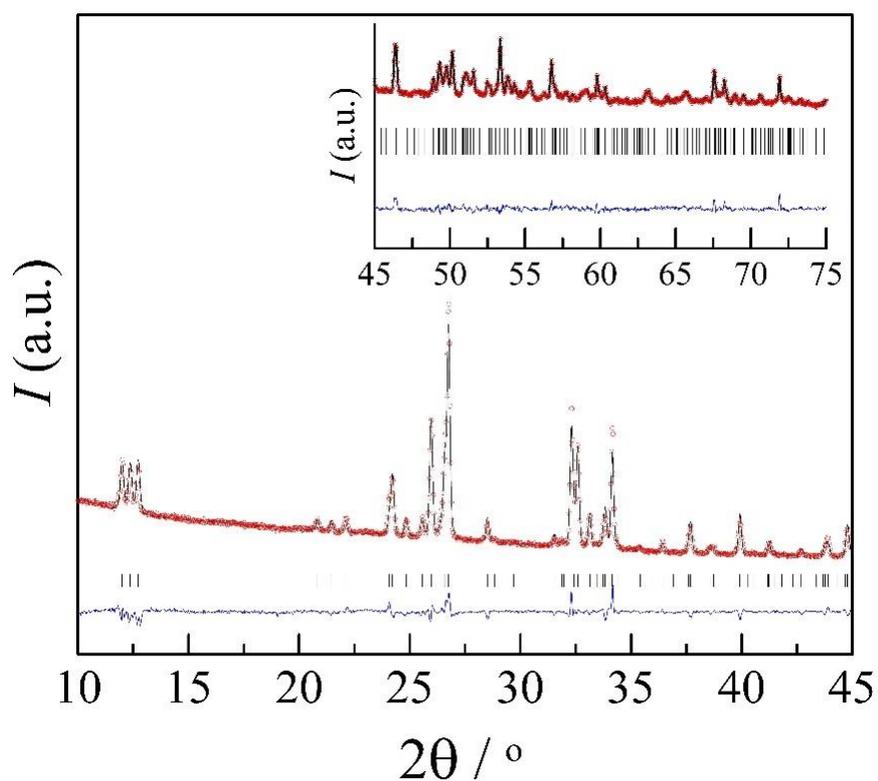

Fig. 7. X-ray diffraction pattern and analysis for TCMX crystalline powder obtained at 100 K. In the main panel are shown experimental (red circles) and calculated (black line) diffraction patterns along with the difference profile (blue line) and Bragg reflections (vertical sticks) of monoclinic $P2_1/c$ space-group. Inset corresponds to the data in the 2θ range between 45 and 75 °.

In Fig. 8, heating and cooling curves obtained from Differential Scanning Calorimetry are depicted. The melting point was observed at 495.5 K in agreement with the expected value for TCMX, hence confirming the purity of the employed TCMX sample.

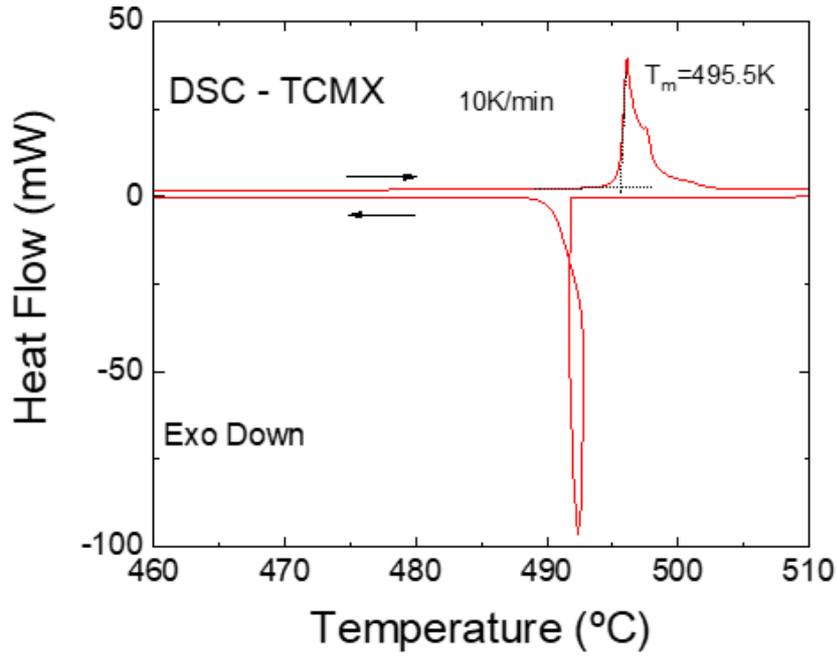

Fig. 8. Differential Scanning Calorimetry (DSC) curves for TCMX crystals. Both upscan and downscan were taken at rate of 10 K/min. The melting point was observed at 495.5 K.

**APPENDIX B: ADDITIONAL SPECIFIC HEAT DATA AND FITS**

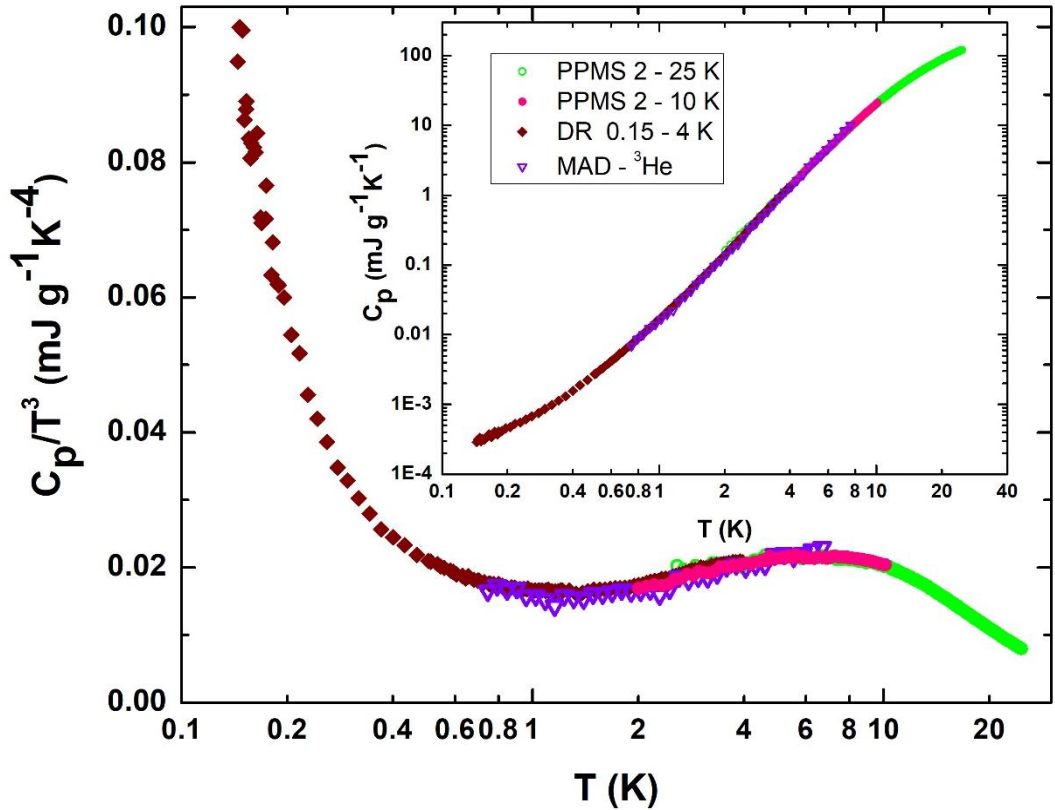

Fig. 9. Specific heat of TCMX crystal in the Debye-reduced $C_p/T^3$ representation for the whole temperature range. Inset: same data in a double logarithmic representation of $C_p(T)$. Different symbols correspond to different experiments, as indicated in the legend.

Fig. 9 shows the whole of specific heat data (after subtraction of the corresponding addenda) of TCMX crystal in an extended temperature range. Different symbols correspond to the use of different experimental setups: $^3$He cryostat in Madrid (down triangles), $^3$He–$^4$He dilution refrigerator (DR) in Wrocław (lozenges) and PPMS system in a standard $^4$He cryostat (circles, two different runs). In the main panel data are presented in the Debye-reduced $C_p/T^3$ representation, whereas in the inset $C_p(T)$ data are depicted in a double logarithmic scale.

In Figs. 10 and 11 we present the lowest temperature data of previous experiments, measured down to 300mK using a $^3$He insert on the PPMS system, with the corresponding SPM fits for the cases of PCNB and $p$-CNB crystals, whose obtained coefficients are presented (Table 1) and discussed in the main text of the article.

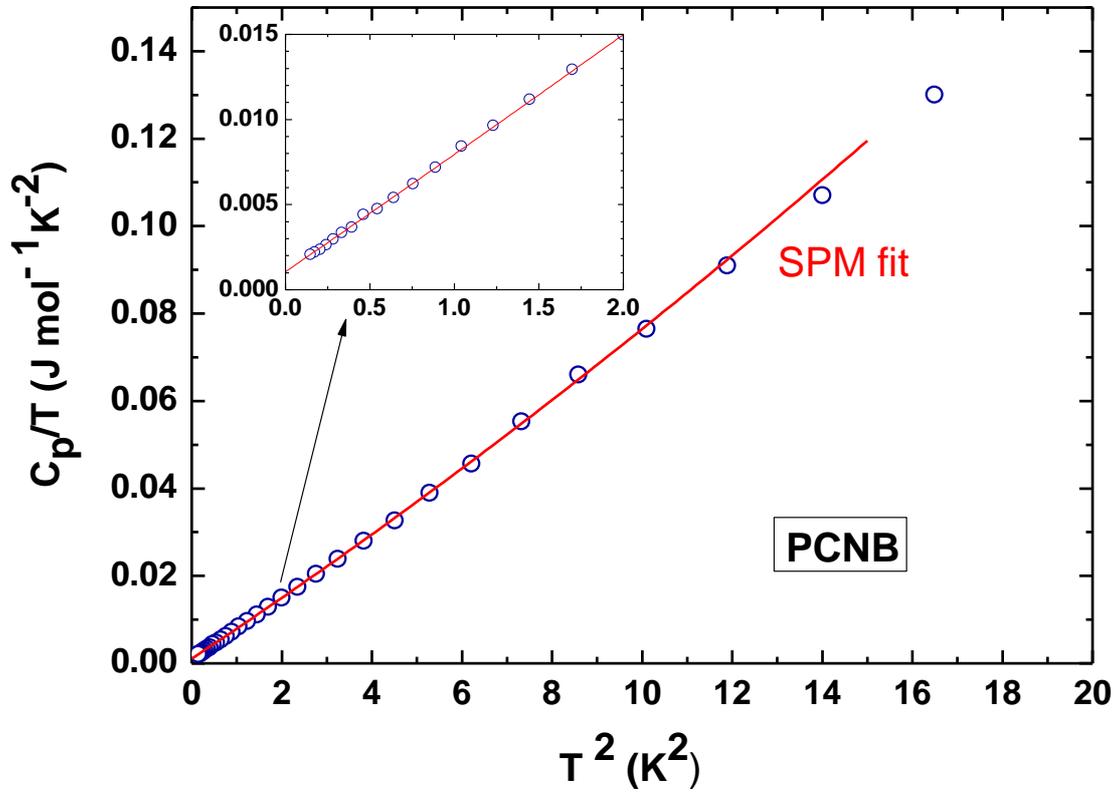

Fig. 10. Specific heat of PCNB crystal at low temperatures in a $C_p/T$ vs $T^2$ representation, from Ref. [18]. The SPM fit is shown by a solid line. The inset amplifies the region at the lowest temperatures.

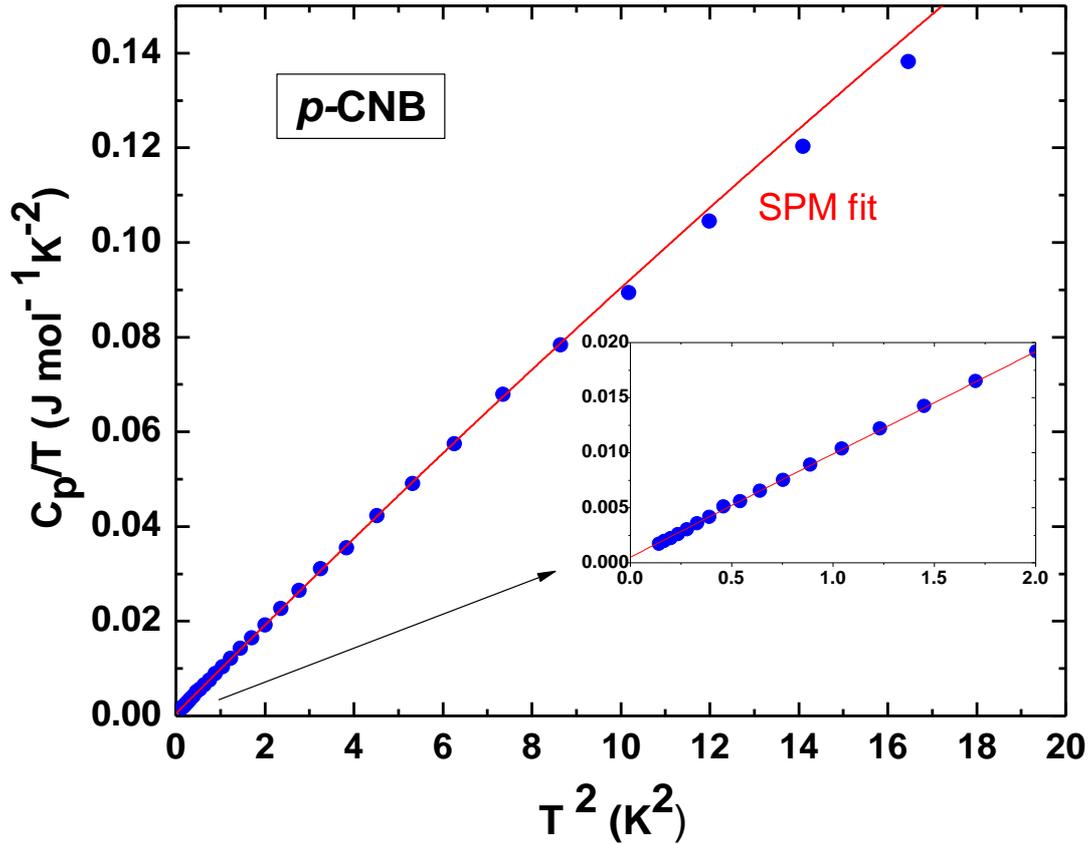

Fig. 11. Specific heat of *p*-CNB disordered crystal at low temperatures in a $C_p/T$ vs $T^2$ representation, using our new data. The SPM fit is shown by a solid line. The inset magnifies the region at the lowest temperatures.

**REFERENCES**

[1] R. C. Zeller and R. O. Pohl. Thermal Conductivity and Specific Heat of Noncrystalline Solids. *Phys. Rev. B* **4**, 2029–2041 (1971).

[2] W. A. Phillips (ed.), *Amorphous Solids: Low Temperature Properties* (Springer, Berlin, 1981).

[3] R. O. Pohl, X. Liu and E. J. Thompson. Low-temperature thermal conductivity and acoustic attenuation in amorphous solids. *Rev. Mod. Phys.* **74**, 991–1013 (2002).

[4] M. A. Ramos (ed.), *Low-temperature thermal and vibrational properties of solids: a half-century of universal ''anomalies'' of glasses* (World Scientific, London-Singapore, 2022).

[5] P. Debye. Zur Theorie der spezifischen Wärmen. *Ann. d. Physik* **39**, 789–807 (1912).

[6] N. W. Ashcroft and N. D. Mermin, *Solid State Physics* (Saunders, Philadelphia, 1976).

[7] C. Kittel, *Introduction to Solid State Physics* (8[th] edn., Wiley, USA, 2005).

[8] E. Bonjour, R. Calemczuk, R. Lagnier, and B. Salce. Low temperature thermal prop-

erties of cyclohexanol: a glassy crystal system. *J. Phys. Col. (France)* **42**, C6-63–65 (1981).

[9] J. J. De Yoreo, W. Knaak, M. Meissner, and R. O. Pohl. Low-temperature properties of Crystalline $(KBr)_{1-x}(KCN)_x$. *Phys. Rev. B* **34**, 8828–8842 (1986).

[10] M. A. Ramos, S. Vieira, F. J. Bermejo, J. Dawidowski, H. E. Fischer, H. Schober, M. A. González, C. K. Loong, and D. L. Price. Quantitative assessment of the effects of orientational and positional disorder on glassy dynamics. *Phys. Rev. Lett.* **78**, 82–85 (1997).

[11] C. Talón, M. A. Ramos, S. Vieira, G. J. Cuello, F. J. Bermejo, A. Criado, M. L. Senent, S. M. Bennington, H. E. Fischer, and H. Schober. Low-temperature specific heat and glassy dynamics of a polymorphic molecular solid *Phys. Rev. B* 58, 745–755 (1998).

[12] P. Esquinazi (ed.), *Tunnelling Systems in Amorphous and Crystalline Solids*, (Springer, Berlin, 1998).

[13] C. Talón, M. A. Ramos, and S. Vieira. Low-temperature specific heat of amorphous, orientational glass, and crystal phases of ethanol. *Phys. Rev. B* **66**, 012201 (2002).

[14] G. A. Vdovichenko, A. I. Krivchikov, O. A. Korolyuk, J. Ll. Tamarit, L. C. Pardo, M. Rovira-Esteva, F. J. Bermejo, M. Hassaine and M. A. Ramos. Thermal properties of halogen-ethane glassy crystals: Effects of orientational disorder and the role of internal molecular degrees of freedom. *J. Chem. Phys.* 143, 084510 (2015).

[15] M. A. Ramos, *Low-Temperature Specific Heat of Glasses and Disordered Crystals,* in *Low-Temperature and Vibrational Properties of Disordered Solids: A Half-Century of "Anomalies" of Glasses*, edited by M. A. Ramos (World Scientific, London/Singapore, 2022), Chapter 2.

[16] A. I. Krivchikov and A. Jeżowski, *Thermal Conductivity of Glasses and Disordered Crystals,* in *Low-Temperature and Vibrational Properties of Disordered Solids: A Half-Century of "Anomalies" of Glasses*, edited by M. A. Ramos (World Scientific, London/Singapore, 2022), Chapter 3.

[17] K. Saito, H. Kobayashi, Y. Miyazaki, and M. Sorai. Anomalous lattice heat capacity of orientationally glassy crystal of *p*-chloronitrobenzene at low temperatures. *Solid State Commun.* 118, 611–614 (2001).

[18] J. F. Gebbia, M. A. Ramos, D. Szewczyk, A. Jeżowski, A. I. Krivchikov, Y. V. Horbatenko, T. Guidi, F. J. Bermejo, and J. Ll. Tamarit. Glassy Anomalies in the Low-Temperature Thermal Properties of a Minimally Disordered Crystalline Solid. *Phys. Rev. Lett.* 119, 215506 (2017).

[19] M. Moratalla, J. F. Gebbia, M. A. Ramos, L. C. Pardo, S. Mukhopadhyay, S. Rudić, F. Fernandez-Alonso, F. J. Bermejo, and J. Ll. Tamarit. Emergence of glassy features in halomethane crystals. *Phys. Rev. B* **99**, 024301 (2019).

[20] A. I. Krivchikov, A. Jeżowski, D. Szewczyk, O. A. Korolyuk, O. O. Romantsova, L. M. Buravtseva, C. Cazorla, and J. Ll. Tamarit. Role of Optical Phonons and

Anharmonicity in the Appearance of the Heat Capacity Boson Peak-like Anomaly in Fully Ordered Molecular Crystals. *J. Phys. Chem. Lett.*, **13**, 5061–5067 (2022).

[21] Y. Miyazaki, M. Nakano, A. I. Krivchikov, O. A. Koroyuk, J. F. Gebbia, C. Cazorla and J. Ll Tamarit. Low-Temperature Heat Capacity Anomalies in Ordered and Disordered Phases of Normal and Deuterated Thiophene. *J. Phys. Chem. Lett.* **12**, 2112–2117 (2021).

[22] D. Szewczyk, J. F. Gebbia, A. Jeżowski, A. I. Krivchikov, T. Guidi, C. Cazorla and J. Ll Tamarit. Heat capacity anomalies of the molecular crystal 1-fluoro-adamantane at low temperaturas. *Sci. Rep.* **11**, 18640 (2021).

[23] S. Hunklinger and W. Arnold, *Ultrasonic Properties of Glasses at Low Temperatures*, in *Physical Acoustics*, edited by W. P. Mason and R. N. Thurston (Academic, New York, 1976), Vol. 12, pp. 155–215.

[24] W. A. Phillips. Two-level states in glasses. *Rep. Prog. Phys.* **50**, 1657–1708 (1987).

[25] U. Buchenau, G. D'Angelo, G. Carini, X. Liu, and M. A. Ramos. Sound absorption in glasses. *Reviews in Physics* **9**, 100078 (2022).

[26] W. A. Phillips. Tunneling states in amorphous solids. *J. Low Temp. Phys.* **7**, 351–360 (1972).

[27] P. W. Anderson, B. I. Halperin, and C. M. Varma. Anomalous thermal properties of glasses and spin glasses. *Philos. Mag.* **25**, 1–9 (1972).

[28] U. Buchenau, N. Nücker and A. J. Dianoux. Neutron scattering study of the low-frequency vibrations in vitreous silica. *Phys. Rev. Lett.* **53**, 2316–2319 (1984).

[29] U. Buchenau, M. Prager, N. Nücker, A. J. Dianoux, N. Ahmad and W. A. Phillips. Low frequency modes in vitreous silica. *Phys. Rev. B* **34**, 5665–5673 (1986).

[30] V. K. Malinovsky, V. N. Novikov, P. P. Parshin, A. P. Sokolov and M. G. Zemlyanov. Universal form of the low-energy (2 to 10 meV) vibrational spectrum of glasses. *Europhys. Lett.* **11**, 43–47 (1990).

[31] M. A. Il'in, V. G. Karpov and D. A. Parshin. Parameters of soft potentials in glasses. *Sov. Phys. JETP* **65**, 165–174 (1987).

[32] U. Buchenau, Yu.M. Galperin, V. L. Gurevich, D. A. Parshin, M. A. Ramos, and H. R. Schober. Interactions of soft modes and sound waves in glasses. *Phys. Rev. B* **46**, 2798–2808 (1992).

[33] T. S. Grigera, V. Martín-Mayor, G. Parisi, and P. Verrocchio. Phonon interpretation of the 'boson peak' in supercooled liquids. *Nature (London)* **422**, 289–292 (2003).

[34] V. L. Gurevich, D. A. Parshin, and H. R. Schober. Anharmonicity, vibrational instability, and the Boson peak in glasses. *Phys. Rev. B* **67**, 094203 (2003).

[35] D. A. Parshin, H. R. Schober, and V. L. Gurevich. Vibrational instability, two-level systems, and the Boson peak in glasses. *Phys. Rev. B* **76**, 064206 (2007).

[36] W. Schirmacher. Thermal conductivity of glassy materials and the "boson peak". *Europhys. Lett*. **73**, 892–898 (2006).

[37] W. Schirmacher, G. Ruocco, and T. Scopigno. Acoustic Attenuation in Glasses and its Relation with the Boson Peak. *Phys. Rev. Lett*. **98**, 025501 (2007).

[38] V. Lubchenko and P. G. Wolynes. The origin of the boson peak and thermal conductivity plateau in low-temperature glasses. *PNAS* **100**, 1515–1518 (2003).

[39] A. I. Chumakov, G. Monaco, A. Monaco, W. A. Crichton, A. Bosak, R. Rüffer, A. Meyer, F. Kargl, L. Comez, D. Fioretto, H. Giefers, S. Roitsch, G. Wortmann, M. H. Manghnani, A. Hushur, Q. Williams, J. Balogh, K. Parliński, P. Jochym, and P. Piekarz. Equivalence of the Boson Peak in Glasses to the Transverse Acoustic van Hove Singularity in Crystals. *Phys. Rev. Lett.* **106**, 225501 (2011).

[40] M. Baggioli and A. Zaccone. Universal origin of boson peak vibrational anomalies in ordered crystals and in amorphous materials. *Phys. Rev. Lett.* **122**, 145501 (2019).

[41] E. Lerner and E. Bouchbinder. Low-energy quasilocalized excitations in structural glasses. *J. Chem. Phys.* **155**, 200901 (2021).

[42] E. Lerner and E. Bouchbinder. Disordered Crystals Reveal Soft Quasilocalized Glassy Excitations. *Phys. Rev. Lett.* **129**, 095501 (2022)

[43] E. Lerner and E. Bouchbinder. Boson-peak vibrational modes in glasses feature hybridized phononic and quasilocalized excitations. *J. Chem. Phys.* **158**, 194503 (2023).

[44] A. J. Leggett. Amorphous materials at low temperatures: why are they so similar? *Phys. B* **169**, 322–327 (1991).

[45] H. M. Carruzzo and C. C. Yu. Why Phonon Scattering in Glasses is Universally Small at Low Temperatures. *Phys. Rev. Lett.* **124**, 075902 (2020).

[46] For a recent review of the Soft-Potential Model, see U. Buchenau, *The Soft-Potential Models and Its Extensions,* in *Low-Temperature and Vibrational Properties of Disordered Solids: A Half-Century of "Anomalies" of Glasses*, ed. by M. A. Ramos (World Scientific, London/Singapore, 2022), Chapter 8.

[47] M. A. Ramos. Are the calorimetric and elastic Debye temperatures of glasses really different? *Phil. Mag*. **84**, 1313–1321 (2004).

[48] T. R. Welberry and D. J. Goossens. The interpretation and analysis of diffuse scattering using Monte Carlo simulation methods. *Acta Cryst. A* **64**, 23–32 (2008).

[49] M. Romanini, M. Barrio, S. Capaccioli, R. Macovez, M. D. Ruiz-Martin, and J. Ll. Tamarit. Double Primary Relaxation in a Highly Anisotropic Orientational Glass-Former with Low-Dimensional Disorder. *J. Phys. Chem. C* **120**, 10614–10621 (2016).

[50] T. C. W. Mak and J. Trotter. The crystal structure of *p*-chloronitrobenzene. *Acta Cryst*. **15**, 1078–1080 (1962).

[51] C. A. Meriles, J. F. Schneider, Y. P. Mascarenhas, and A. H. Brunetti. X-ray diffraction study of polycrystalline p-chloronitrobenzene. *J. Appl. Cryst.* **33**, 71−81 (2000).

[52] L. H. Thomas, J. M. Cole, and C. C. Wilson. Orientational disorder in 4-chloronitrobenzene *Acta Cryst. C* **64**, 296−302 (2008).

[53] T. Bräuniger, R. Poupko, Z. Luz, D. Reichert, H. Zimmermann, H. Schmitt, and U. Haeberlen. Orientational disorder in 1,2,3-trichloro-4,5,6-trimethylbenzene. A single crystal deuterium NMR study of the site populations and dynamics. *Phys. Chem. Chem. Phys.* **3**, 1891−1903 (2001).

[54] E. Pérez-Enciso and M. A. Ramos. Low-temperature calorimetry on molecular glasses and crystals. *Thermochimica Acta* **461**, 50−56 (2007).

[55] T. Pérez-Castañeda, J. Azpeitia, J. Hanko, A. Fente, H. Suderow, and M. A. Ramos. Low-Temperature Specific Heat of Graphite and CeSb$_2$: Validation of a Quasi-adiabatic Continuous Method. *J. Low Temp. Phys.* **173**, 4−20 (2013).

[56] Y. Tozuka, Y. Yamamura, K. Saito, and M. Sorai. Thermodynamic study of a phase transition between the ordered and disordered phases and orientational disorder in crystalline *p*-chloronitrobenzene. *J. Chem. Phys.* **112**, 2355−2360 (2000).

[57] T. Pérez-Castañeda, C. Rodríguez-Tinoco, J. Rodríguez-Viejo, and M. A. Ramos. Suppression of tunneling two-level systems in ultrastable glasses of indomethacin. *PNAS* **111**, 11275−11280 (2014).

[58] J. A. Katerberg and A. C. Anderson. Low-temperature specific heat, thermal conductivity, and ultrasonic velocity of glassy carbons. *J. Low Temp. Phys.* **30**, 739−745 (1978).

[59] J. Etrillard, J. C. Lasjaunias, B. Toudic, and H. Cailleau. Low-frequency excitations in incommensurate biphenyl as studied by very low-temperature specific heat. *Europhys. Lett.* **38**, 347−352 (1997).

[60] R. Li, H. Zheng, Y. Hua, R. Wei, Z. Tan, and Q. Shi. Low temperature heat capacity study of Zn, Cd and Mn based coordination compounds synthesized using phenanthroline

[61] R. Takehara, N. Kubo, M. Ryu, S.Kitani, S. Imajo, Y. Shoji, H. Kawaji, J. Morikawa, and T. Fukushima. Insights into Thermal Transport through Molecular *π*-Stacking. *J. Am. Chem. Soc.* **145**, 22115−22121 (2023).

[62] M. Romanini, I. B. Rietveld, M. Barrio, Ph. Negrier, D. Mondieig, R. Macovez, R. Céolin, and J.-Ll. Tamarit. Uniaxial Negative Thermal Expansion in Polymorphic 2-Bromobenzophenone, Due to Aromatic Interactions? *Cryst. Growth Des.* **21**, 2167−2175 (2021).

[63] N. Mahé, B. Nicolaï, H. Allouchi, M. Barrio, B. Do, R. Céolin, J.-Ll. Tamarit, and I. B. Rietveld. Crystal Structure and Solid-State Properties of 3,4-Diaminopyridine Dihydrogen Phosphate and Their Comparison with Other Diaminopyridine Salts. *Cryst. Growth Des.* **13**, 3028-3035 (2013).

[64] B. Nicolaï, I. B. Rietveld, M. Barrio, N. Mahé, J.-Ll. Tamarit, and R. Céolin. Uniaxial negative thermal expansion in crystals of tienoxolol. *Struct. Chem.* **24**, 279–283 (2013).

[65] C. Červinka, V. Štejfa, V. Pokorný, P. Touš, and K. Růžička. Orientational Disorder in Crystalline Disubstituted Benzenes and Its Implications for Sublimation and Polymorphism. *Cryst. Growth Des.* **23**, 9011-9024 (2023).

[66] R. N. Goldberg, J. Schliesser, A. Mittal, S. R. Decker, A.F.L. Santos, V. L. Freitas, and D. K. Johnson. A thermodynamic investigation of the cellulose allomorphs: Cellulose (am), cellulose Iβ (cr), cellulose II (cr), and cellulose III (cr). *The J. Chem. Thermodyn.* **81**, 184-226 (2015).

[67] M. Mohr, J. Maultzsch, E. Dobardžić, S. Reich, I. Milošević, M. Damnjanović, A. Bosak, M. Krisch, and C. Thomsen. Phonon dispersion of graphite by inelastic x-ray scattering. *Phys. Rev. B* **76**, 035439 (2007).

[68] M. Popovic, G. B. Stenning, A. Göttlein, and M. Minceva. Elemental composition, heat capacity from 2 to 300 K and derived thermodynamic functions of 5 microorganism species. *J. biotech.* **331**, 99-107 (2021).

[69] A. Taheri, S. Pisana, and Ch. Veer Singh. Importance of quadratic dispersion in acoustic flexural phonons for thermal transport of two-dimensional materials. *Phys. Rev. B* **103**, 235426 (2021).

[70] L. Landau and E. Lifshitz. *Theory of Elasticity* (Pergamon, Oxford, 1995).

[71] J. Sibik and J. A. Zeitler. Direct measurement of molecular mobility and crystallisation of amorphous pharmaceuticals using terahertz spectroscopy. *Adv. Drug Deliv. Rev.* **100**, 147–157 (2016). and halogenated benzoic acid. *Thermochimica Acta* **670**, 76–86 (2018).

[72] A.V. Granato. Interstitialcy model for condensed matter states of face-centered-cubic metals. *Phys. Rev. Lett.* **68**, 974–977 (1992).

[73] A.V. Granato. Interstitial resonance modes as a source of the boson peak in glasses and liquids. *Physica B* **219&220**, 270–272 (1996).

[74] A.V. Granato. The specific heat of simple liquids. *J. Non Cryst. Solids* **307–310**, 376–386 (2002).

[75] G. Carini, G. D´Angelo, G. Tripodo, and G. A. Saunders. Specific heat and low-energy excitations of samarium phosphate glasses. *Phil. Mag.* **71**, 539–545 (1995).

[76] D.-M. Zhu and H. Chen. Low temperature specific heat and fragility of glasses. *J. Non-Cryst. Solids* **224**, 97–101 (1998).

[77] M. A. Ramos, C. Talón, and S. Vieira. The Boson peak in structural and orientational glasses of simple alcohols: specific heat at low temperatures. *J. Non-Cryst. Solids* **307–310**, 80–86 (2002).

[78] M. Moratalla, M. Rodríguez-López, C. Rodríguez-Tinoco, J. Rodríguez-Viejo, R. J. Jiménez-Riobóo, and M. A. Ramos. Depletion of two-level systems in highly stable glasses with different molecular ordering. *Communications Physics* **6**, 274 (2023).

[79] L. Wang, A. Ninarello, P. Guan, L. Berthier, G. Szamel, and E. Flenner. Low-frequency vibrational modes of stable glasses. *Nature Communications* **10**, 26 (2019).

[80] H. Mizuno, M. Shimada, and A. Ikeda. Anharmonic properties of vibrational excitations in amorphous solids. *Phys. Rev. Res.* **2**, 013215 (2020).